\documentclass[12pt]{iopart}

\usepackage{graphicx}

\usepackage{iopams}  
\usepackage[usenames,dvipsnames]{color}   

\usepackage[utf8]{inputenc}
\usepackage[T1]{fontenc}
\usepackage{verbatim}

\usepackage{amssymb}
\usepackage{amsfonts}

\newcommand{\kB}{\ensuremath{k_{{B}}}}		

\renewcommand{\vec}{\mathbf}

\renewcommand{\vec}[1]{\mathbf{#1}}

\begin{document}

\title[Screened Coulomb potential in a flowing magnetized plasma]{Screened Coulomb potential in a flowing magnetized plasma}

\author{J-P Joost$^1$, P Ludwig$^1$, H K\"ahlert$^1$, C Arran$^2$, and M Bonitz$^1$}

\address{$^1$ Institut f\"ur Theoretische Physik und Astrophysik, Christian-Albrechts-Universit\"at zu Kiel, Leibnizstr. 15, 24118 Kiel, Germany}
\address{$^2$ Emmanuel College, Cambridge, UK}

\ead{ludwig@theo-physik.uni-kiel.de}
\begin{abstract}
The electrostatic potential of a moving dust grain in a complex plasma with magnetized ions is computed using linear response theory, thereby extending our previous work for unmagnetized plasmas~[P. Ludwig \textit{et al.}, New J. Phys. \textbf{14}, 053016 (2012)]. In addition to the magnetic field, our approach accounts for a finite ion temperature as well as ion-neutral collisions. Our recently introduced code \texttt{Kielstream} is used for an efficient calculation of the dust potential. Increasing the magnetization of the ions, we find that the shape of the potential crucially depends on the Mach number $M$. In the regime of subsonic ion flow ($M<1$), a strong magnetization gives rise to a potential distribution that is qualitatively different from the unmagnetized limit, while for $M>1$ the magnetic field effectively suppresses the plasma wakefield.
\end{abstract}

\maketitle

\section{Introduction}
Plasma wakes are a fascinating collective phenomenon, which can give rise to attraction of like charged particles in multi-component plasmas. In complex plasmas~\cite{2010BonitzReview}, dynamical screening and wake effects have been investigated in a large number of studies, including experimental~\cite{komo2000,savl2009,krsc2010,springer2014partA} as well as theoretical~\cite{lampe2000,chkh2010,dewar2012,lumi2012,komp2014} work.\footnote{For a more extensive list of earlier work see~\cite{blca2012,springer2014}.}
Computational approaches include first-principle molecular dynamics simulations~\cite{mavl2000,vlma2003,grsu2013}, fluid codes~\cite{mego95,schwabe2013} and particle-in-cell (PIC) simulations, e.g.~\cite{schneider2010,hupa2010,hutch2011,mibl2012}.

With the recent availability of superconducting magnets in several laboratories~\cite{schwabe2011,thomas2012}, the focus has shifted towards plasmas where the effect of strongly magnetized ions on the dynamically screened dust potential can be studied in detail~\cite{carst2012,2014jung}. Theoretically, the screening of a test charge in a magnetized plasma has been studied in various publications~\cite{shsa1996,sana2000,nasa2001a,nasa2001b,nasa2001c,na2001,shna2001,nani2002,nina2003,sato2003,sash2003,sari2004,sali2007,avsh2012,bhda2012,bhda2013}, but many details remain unclear. These studies typically found an oscillatory wake pattern, as in the unmagnetized case, but with a magnetic field-dependent oscillation period and amplitude. 
However, the predicted trends of an increasing magnetic induction on the amplitude and wave length are contradictory.
For an increasing external magnetic field parallel to the ion flow direction, amplification~\cite{shsa1996, sana2000, shna2001, nina2003, sato2003, sash2003, bhda2012, bhda2013} as well as damping~\cite{nasa2001a,na2001,nani2002,sato2003} of the wake oscillations has been reported.
Moreover, the ions were mostly treated within a fluid approach and, consequently, kinetic effects were neglected. For typical experimental parameters ($T_e/T_i\approx 100$), however, their influence can be substantial. In particular, it has been shown~\cite{lampe2005ieee,hutch2012,lumi2012} that Landau damping can significantly damp the wake oscillations in unmagnetized plasmas.
Therefore, similar effects are expected for magnetized plasmas as well. Moreover, there are additional contributions to the dielectric function related to ion Bernstein modes~\cite{bernstein1958}, which propagate perpendicular to the magnetic field~\cite{pahu2007,pahu2011}.

The main goal of this paper is a systematic description of the topology of the wake structure in real space over broad parameter ranges: (i) from the subsonic to the supersonic regime of ion streaming, and (ii) weak to strong ion magnetization where the field is aligned with the flow.
The calculations are based on a recently developed high performance linear response code that allows for an effective computation of the potential on very large grids~\cite{springer2014partB}.


\section{Dielectric function approach}
The linear response of a partially ionized plasma to an external perturbation, such as a moving dust particle with charge $Q$, can be calculated from the longitudinal dielectric function (DF)~\cite{lampe2005ieee,ludwigquantum}, $\varepsilon(\vec k,\omega)=1+\chi_e+\chi_i$, which contains contributions from the electrons and ions (susceptibilities $\chi_{e,i}$). The neutral gas does not enter the DF directly but can modify the ion response considerably due to ion-neutral collisions with frequency~$\tilde\nu_i$. The real space potential at the point $\vec R=\vec r-\vec u_{d}\, t$ is given by~\cite{ludwigquantum}
\begin{equation}\label{eqn:potential}
 \Phi(\vec R)=\int \frac{d^3 k}{2\pi^2}\frac{Q}{k^2} \frac{e^{i\vec k\cdot \vec R}}{\varepsilon(\vec k,\vec k\cdot \vec u_{d})}.
\end{equation}
Equation~(\ref{eqn:potential}) can account for streaming ions (via the ion velocity distribution) as well as moving dust particles (via the argument $\omega=\vec k\cdot \vec u_{d}$ of the DF, where $\vec u_{d}$ is the velocity of the dust particle).

A kinetic study of waves in a magnetized Maxwellian plasma was first conducted by Bernstein~\cite{bernstein1958}. The derivation of the associated dielectric function can be found in classical textbooks~\cite{alexandrov2013,bellan2006}. As in our previous work~\cite{lumi2012}, we use a BGK collision term to account for ion-neutral collisions (collision frequency $\tilde\nu_{i}$). The ion susceptibility can then be written as~\cite{alexandrov2013}
 \begin{equation}\label{eqn:ionDF}
    \chi_i(\vec k,\omega)=\frac{1}{k^2 \lambda_{Di}^2}\frac{1+\sum_{n=-\infty}^\infty \frac{\omega+i\tilde\nu_i}{\omega+i\tilde\nu_i-n\omega_{ci}}e^{-\eta_i}I_n(\eta_i)\, \zeta_{i,n} Z(\zeta_{i,n})  }{1+\sum_{n=-\infty}^\infty \frac{i\tilde\nu_i}{\omega+i\tilde\nu_{i}-n\omega_{ci}}e^{-\eta_i}I_n(\eta_i) \, \zeta_{i,n} Z(\zeta_{i,n})},
 \end{equation}
where $I_n(\eta_i)$ is the modified Bessel function of the first kind, $Z(\zeta_{i,n})$ the plasma dispersion function~\cite{zaghloul}, $\eta_{i}=k_\perp^2 v_{th,i}^2/\omega_{ci}^2$, and $\zeta_{{i,n}}=(\omega+i\tilde\nu_{i}-n\omega_{ci})/(\sqrt{2}\,|k_z|v_{th,i})$~\footnote{We note that on page 133 in \cite{alexandrov2013} the modulus of the wave number is missing.}. 
The ion thermal velocity and the ion cyclotron frequency are given by $v_{th,i}=(\kB T_{i}/m_{i})^{1/2}$ and 
$\omega_{ci}=q_{i} B/m_{i}$, 
respectively. The magnetic field is chosen parallel to the ion streaming direction.
The electron response is treated in the static approximation, i.e., $\chi_e=(k\lambda_{De})^{-2}$,
where $\lambda_{De}=\sqrt{\varepsilon_0 k_B T_e / (n_e q_e^2)}$ is the electron Debye length.

In the sheath region of an rf discharge, the ions stream past the dust grain with a mean velocity $\vec u_{i}$, and the latter can be considered at rest, $\vec u_{d}=0$. On the other hand, in the rest frame of the ions, where $\vec u_{i}'=0$, the dust particles move with a velocity $\vec u_{d}'=-\vec u_{i}$. Thus, it appears that we can equally evaluate Eq.~(\ref{eqn:potential}) in the rest frame of the ions, where the ion susceptibility [Eq.~(\ref{eqn:ionDF})] is well known. However, this apparent symmetry is broken by the presence of the neutral gas, see Ref.~\cite{ivlev2005}. Equation~(\ref{eqn:ionDF}) is only valid for ions in thermal equilibrium. It does not account for the fact that the ions are being accelerated by the sheath electric field, which causes (i) a net drift of the ions with respect to the gas and (ii) deviations of the ion distribution function from a Maxwellian. Nevertheless, the calculation outlined above may serve as a starting point to explore the effect of a magnetic field on the screening potential within a kinetic framework.


 
%
%

\section{Numerical implementation}

The computation of the dynamically screened ion (wake) potential in real space is based on a numerical three-dimensional discrete Fourier transformation (3D DFT) on a relatively large grid with resolutions from $1024 \times 1024 \times  4096$ up to $4096 \times 4096 \times 16384$.~\footnote{In order to avoid pseudo-periodical effects, the range in real space must be chosen in all dimensions with proper size, whereby more grid points are required in the streaming direction.} In order to handle 3D grids of this size, the previously introduced high performance linear response program \texttt{Kielstream} is used~\cite{springer2014partB}. \texttt{Kielstream} was developed in \texttt{C++} to calculate the screened plasma potential for the unmagnetized case. The program is optimized for memory efficiency and achieves high performance by parallelization using the \texttt{openMP} library and by exploiting the radial symmetry of the problem. In addition it uses the \texttt{libcerf} library~\cite{zaghloul,libcerf} to reliably evaluate the plasma dispersion function and the \texttt{fftw3} library~\cite{fftw3} to efficiently perform the Fourier transformation. The modified Bessel function is evaluated using the \texttt{GNU Scientific Library (GSL)}~\cite{gsl}.

\begin{figure}[t]
\includegraphics[width=0.9\textwidth]{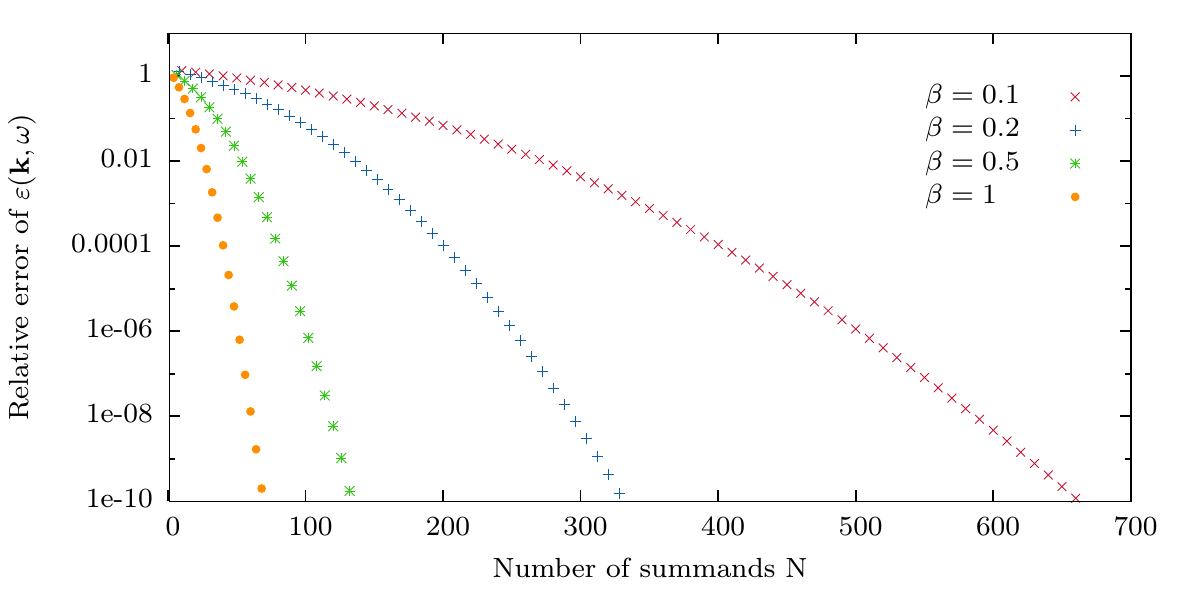}\hspace{2pc}%
\begin{minipage}[b]{\textwidth}\caption{\label{pic:convergence}
Relative error in computing the real part of the dielectric function $\varepsilon(\vec k,\omega)=1+\chi_e+\chi_i$ depending on the total number of evaluated summands of the infinite sum occurring in equation~(\ref{eqn:ionDF}).
Approaching weak magnetization, for each grid point in k-space several hundreds of terms are required to ensure convergence. 
Parameters: $k=k_z=k_\perp=50\lambda_{De}^{-1}$ and $M=1$, $T_e/T_i=100$, $\nu_i=\tilde\nu_i/\omega_p=0.003$. 
The magnetization is described by the ratio of the ion cyclotron and the ion plasma frequency, $\beta=\omega_{ci}/\omega_p$, see section~\ref{sec:results}.}
\end{minipage}
\end{figure}

For the calculation of the screened potential we have to carry out two steps: (i)~the population of the grid and (ii) the execution of the 3D DFT. 
Modifications to \texttt{Kielstream} in the context of this work are related to the first part, as the dielectric function for the magnetized ions, Equation~(\ref{eqn:ionDF}), has to be implemented. Compared to the unmagnetized case, the key issue, from the numerical point of view, is the appearance of the infinite sum involving expensive multiple evaluations of the Bessel and the complex plasma dispersion functions for every specific value of $\vec k$. 
The number of elements that must be summed up to ensure convergence of the real and imaginary part crucially depends on the magnetization $\beta=\omega_{ci}/\omega_p$~\cite{alexandrov2013}.\footnote{
Note that one has to ensure the convergence of the infinite sums in equation~(\ref{eqn:ionDF}) for every point on the 3D grid since
the convergence explicitly depends also on $\vec k$.}
Especially for small magnetization, $\beta \rightarrow 0$, the sum converges very slowly. Then the complicated special functions of the
summand have to be evaluated up to several hundred times for every point on the 3D grid which greatly increases the complexity of the problem compared to the unmagnetized case, see figure~\ref{pic:convergence}.
We note that the plasma dispersion function as well as the Bessel function reveal specific invariances that can be exploited for numerical optimizations. 
That is, in contrast to the unmagnetized case, the plasma dispersion function does not depend on $k_\perp$ 
while on the other hand the Bessel function does not depend on $k_z$. Using large tables for the required number of summands at the characteristic $k$-points, the numerical effort of evaluating the special functions and therefore the time for the population of the grid can be substantially reduced.
Without optimizations the time used for the population of the 3D grid dominates the overall computation time.






\section{Results} \label{sec:results}
The plasma wakefield depends on four dimensionless parameters: the Mach number $M=u_d/c_s$, the electron-to-ion temperature ratio $T_e/T_i$, the ion-neutral collision frequency $\nu_i=\tilde\nu_i/\omega_p$, as well as the magnetization of ions $\beta=\omega_{ci}/\omega_{p}$, with the plasma frequency $\omega_p=\sqrt{n_i q_i^2/(\varepsilon_0 m_i)}$. Here, the Mach number is defined as the ratio of the ion streaming velocity $u_i$ and the ion sound speed, $c_s=\sqrt{k_B T_e/m_i}$. Without loss of generality, we consider a grain charge of $Q_d=-10^4e_0$ and an electron Debye length of $\lambda_{De}=0.845$\,mm. Due to the linear response ansatz, the potential can be simply rescaled to any other grain charge of interest.

In the following, we consider a fixed temperature ratio of $T_e/T_i=100$ giving rise to pronounced wakefields. Furthermore, this value is often considered in PIC simulations~\cite{lumi2012,hutch2011}. Two values of the ion-neutral collision frequency are studied: (i) $\nu_{i}=0.003$, which applies to the collisionless case\footnote{We note, that some finite damping is required for numerical reasons, in order to avoid convergence issues giving rise to pseudo-periodical effects.\cite{springer2014}}, and (ii) $\nu_{i}=0.1$, which is representative for typical experimental setups. The Mach number is varied in the range $M=0.33\dots 1.5$ taking into account that for very small ion streaming velocities the linear response approach may not be applicable due to strong dust-plasma interactions. Our main interest is the dependence of the wake potential on the magnetization of the ion plasma background. Therefore, a broad range of magnetic inductions $\beta=0\ldots10$ is considered.
The ion Larmor radius of gyration $r_L=v_{th,i}/\omega_c$ is in units of the electron Debye length given by $r_L/\lambda_{De}=\beta^{-1}\sqrt{T_i/T_e}$, e.g., $r_L/\lambda_{De}=1,0.1,0.01$ for $\beta=0.1,1,10$, respectively.

\subsection{Subsonic regime}

\begin{figure}[p]
\includegraphics[width=0.95\textwidth]{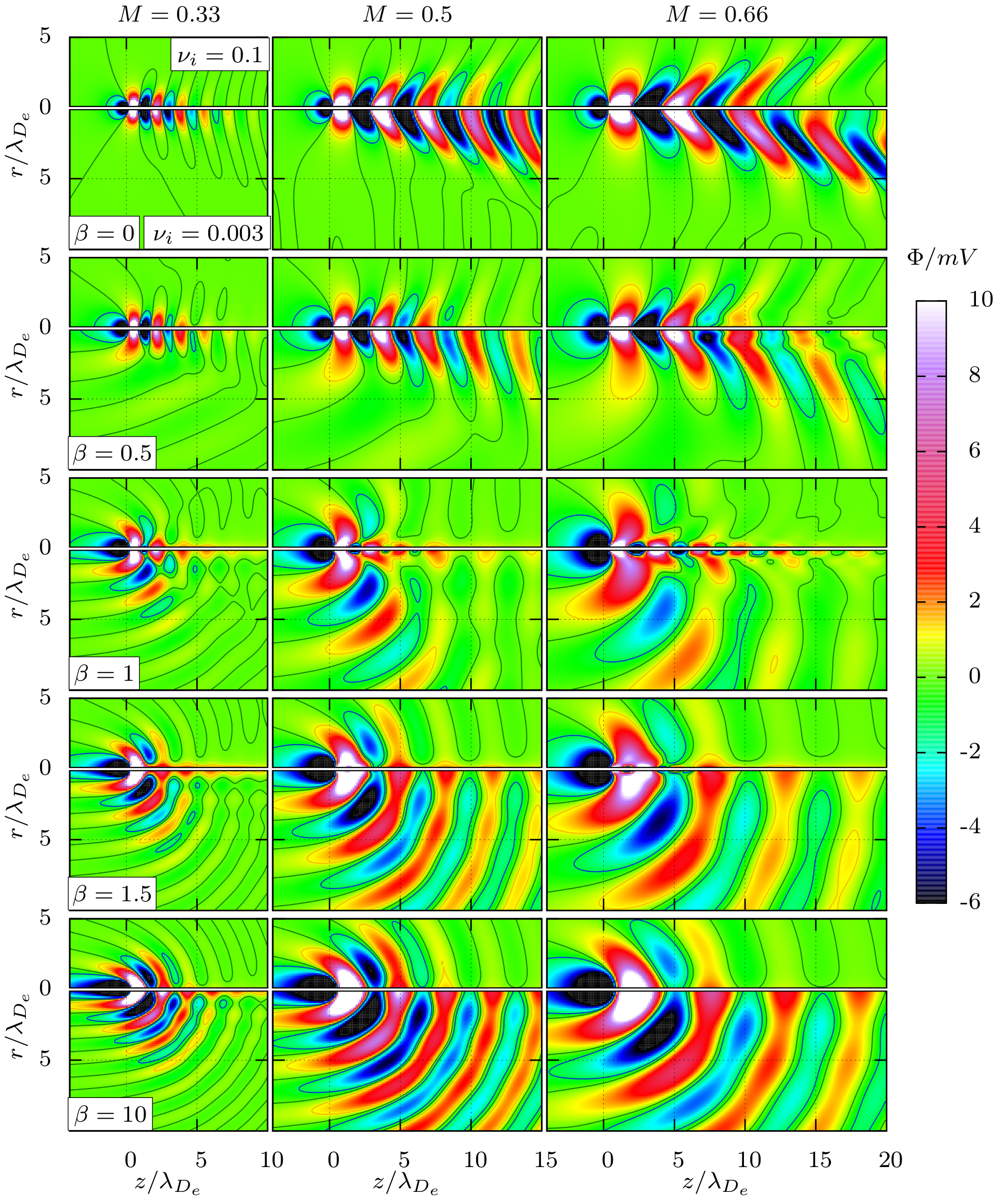}\hspace{2pc}%
\begin{minipage}[b]{\textwidth}\caption{\label{pic:matrix1}
Subsonic regime: Contour plot of plasma potential $\Phi(r)$, see Eq.~(\ref{eqn:potential}),  for three different values of ion streaming velocities  (from left to right: $M=0.33, 0.5$ and $M=0.66$) and five different levels of the external magnetic field (increasing from top to bottom). 
The upper half of each panel shows the case of finite damping ($\nu=0.1$) while the lower one corresponds to the (almost) collisionless case ($\nu=0.003$), where the plasma oscillation are only weakly damped. The dust grain is located at the origin. The ions are streaming from left to right. 
Yellow/red to white (blue to black) colours correspond to positive (negative) potential values. 
Equipotential lines are shown for $-1$\,mV (blue), $0$\,mV (dark green) and $1$\,mV (orange), respectively.}
\end{minipage}
\end{figure}

In figure \ref{pic:matrix1}, we present a contour plot of the dynamically screened dust potential. Note that in the presence of streaming ions the potential has a three-dimensional conical shape. Due to the cylindrical symmetry of the potential, the plot uses cylindrical coordinates with the $z$ axis being aligned with the ion streaming velocity and the magnetic field. As a reference, let us first consider the unmagnetized case, $\beta=0$ (top row), see also Ref.~\cite{lumi2012}. Even for $M=0.33$, strong deviations from the isotropic Yukawa potential are apparent, for both the collisionless plasma, in the lower panel, and when finite damping ($\nu_i=0.1$) is included, in the upper panel. In the direct vicinity of the grain, there is a strong Yukawa-like repulsive potential region. In the streaming direction, we find an attractive potential part right behind the grain which attracts other negatively charged grains downstream. Increasing the Mach number $M$, the potential peaks on the symmetry axis are shifted away from the grain, and the range of the wakefield increases. On the other hand, the angle of the conic wavefronts decreases. For $M=0.66$, we find that the potential peaks break away from the centre axis at large distances, $z \gtrsim 10 \lambda_{De}$. Generally, the extension of the wakefield increases with $M$.

Considering now a finite magnetization, $\beta=0.5$, we find several qualitative and quantitative differences compared to the unmagnetized limit, cf. Fig.~\ref{pic:matrix1} (lower rows). First, screening in streaming direction becomes stronger. This effect is even more pronounced for the collisionless case. In particular, the amplitude of the wakefield is significantly reduced. This is clearly visible, especially in the long tail of the plasma wake. Second, the equipotential lines become strongly bent. This effect becomes more and more pronounced for larger values of $\beta$. Independently of $M$, additional potential peaks appear off the z-axis, as seen for $\beta=1$, while the wake oscillations near the z-axis--and hence the attraction of downstream particles--are strongly reduced. Third, screening becomes weaker with increase of $\beta$ in the upstream direction.

Approaching very strong magnetization, $\beta=10$, the 3D wake structure creates nested half shells around the grain. That is, compared to the unmagnetized case, the direction of the cone-structure is reversed.\footnote{The formation of similar wake pattern behind the grain in the subsonic regime has also been described in Ref.\cite{nina2003}; however, for a magnetic field of $\beta\approx 0.3$.}  This effect is, however, strongly reduced in the collisional case. We note that keeping only the $n=0$ term in the ion susceptibility, Eq.~(\ref{eqn:ionDF}), which corresponds to the limit $\beta\to \infty$ (see also figure~\ref{pic:convergence}), we find that the potential pattern changes only marginally compared to $\beta=10$.

\begin{figure}[t]
\includegraphics[width=0.935\textwidth]{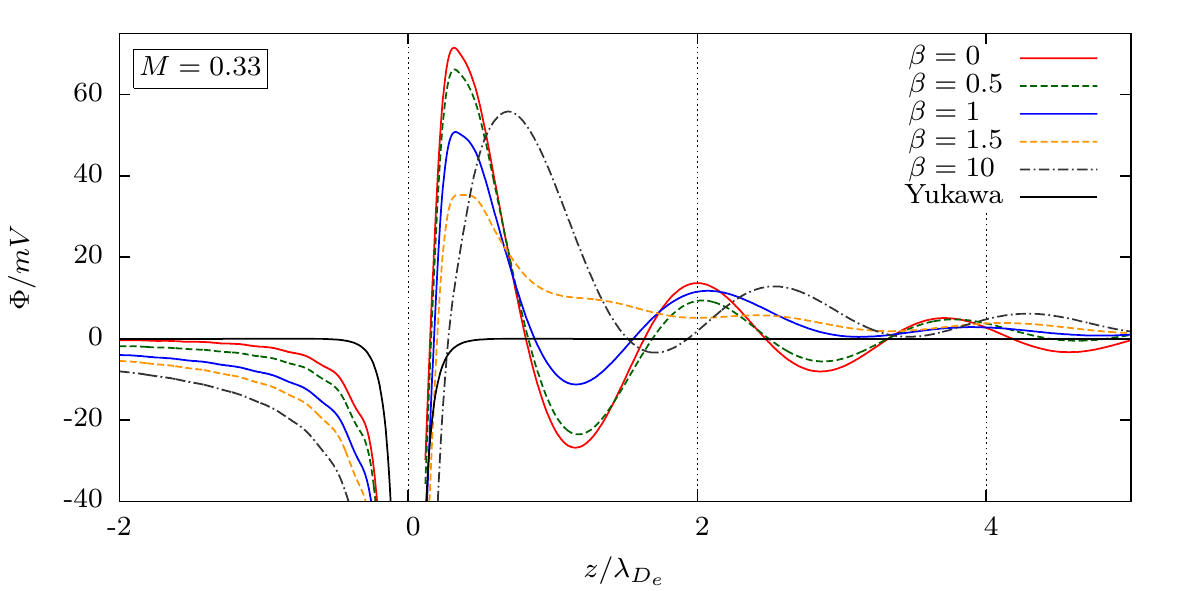}\hspace{2pc}%
\begin{minipage}[b]{\textwidth}\caption{\label{pic:axial033}
Potential cuts through the grain ($r = 0$) along the flow direction for different magnetic inductions $\beta$ in the subsonic regime $M=0.33$ ($\nu=0.003$, $T_e/T_i=100$). 
Also shown is the Yukawa potential for the corresponding static case (black solid line, $M=0$).
Cf. figures~\ref{pic:axial066}, \ref{pic:axial1}, and \ref{pic:axial15}.}
\end{minipage}
\end{figure}

\begin{figure}[t]
\includegraphics[width=0.935\textwidth]{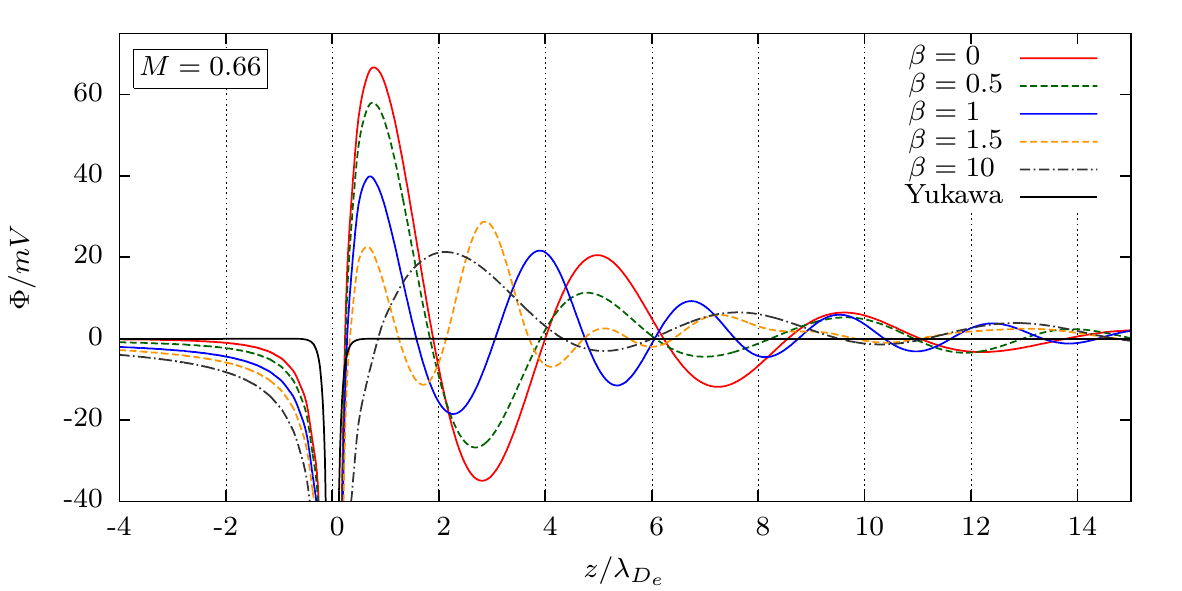}\hspace{2pc}%
\begin{minipage}[b]{\textwidth}\caption{\label{pic:axial066}
Same as figure \ref{pic:axial033} but for $M=0.66$. Note the different scaling of the $z$-axis.
}
\end{minipage}
\end{figure}

In order to point out the unrestricted effect of magnetization, the collision frequency is set to $\nu_i = 0.003$ in the following, i.e., the (almost) collisionless case is considered.
The potential profile along the z-axis for $M=0.33$ is shown in figure~\ref{pic:axial033}. As mentioned above, there is a clear trend that screening decreases with increasing magnetization in the upstream direction. While the potential in this direction is purely monotonic, there are strong oscillations in the potential downstream of the grain. In particular, the position of the first peak is found to be almost independent of $\beta$~\cite{sana2000}. However, in the limiting case of large magnetization, the peak is shifted away from the grain, which can be attributed to the fact that the off-axis extrema reconnect, see figure~\ref{pic:matrix1}. 
Another notable result is that for an intermediate value of $\beta=1.5$, there is no (negative) potential minimum in the flow direction at all. Minima are observed only off the $z$-axis.

This anomaly, however, disappears at a higher ion streaming velocity, $M=0.66$, see figure~\ref{pic:axial066}. It is immediately clear that the wake structure is much more extended since the wavelength of the wake oscillations increase with $M$. Typically, the peaks are slightly shifted towards the grain as $\beta$ is increased~\cite{bhda2012}. Again, a strong deviation from this trend is observed for the limiting case of large $\beta$, where the first potential peak is much broader and located at about twice the distance from the grain compared to the unmagnetized limit.

In contrast to earlier predictions for subsonic ion flow~\cite{shsa1996, sana2000, shna2001, nina2003,bhda2012}, the amplitude of the oscillatory wake potential generally decreases with increasing magnetic induction.\footnote{We note a deviation from this general trend for $M=0.33$ in the limiting case $\beta \rightarrow \infty$; while the amplitude of the trailing peak is lower than in unmagnetized case, it is significantly larger than for intermediate values of $\beta$ due to the reconnection of lateral extrema.}

\subsection{Supersonic regime}

\begin{figure}[t]
\includegraphics[width=1.0\textwidth]{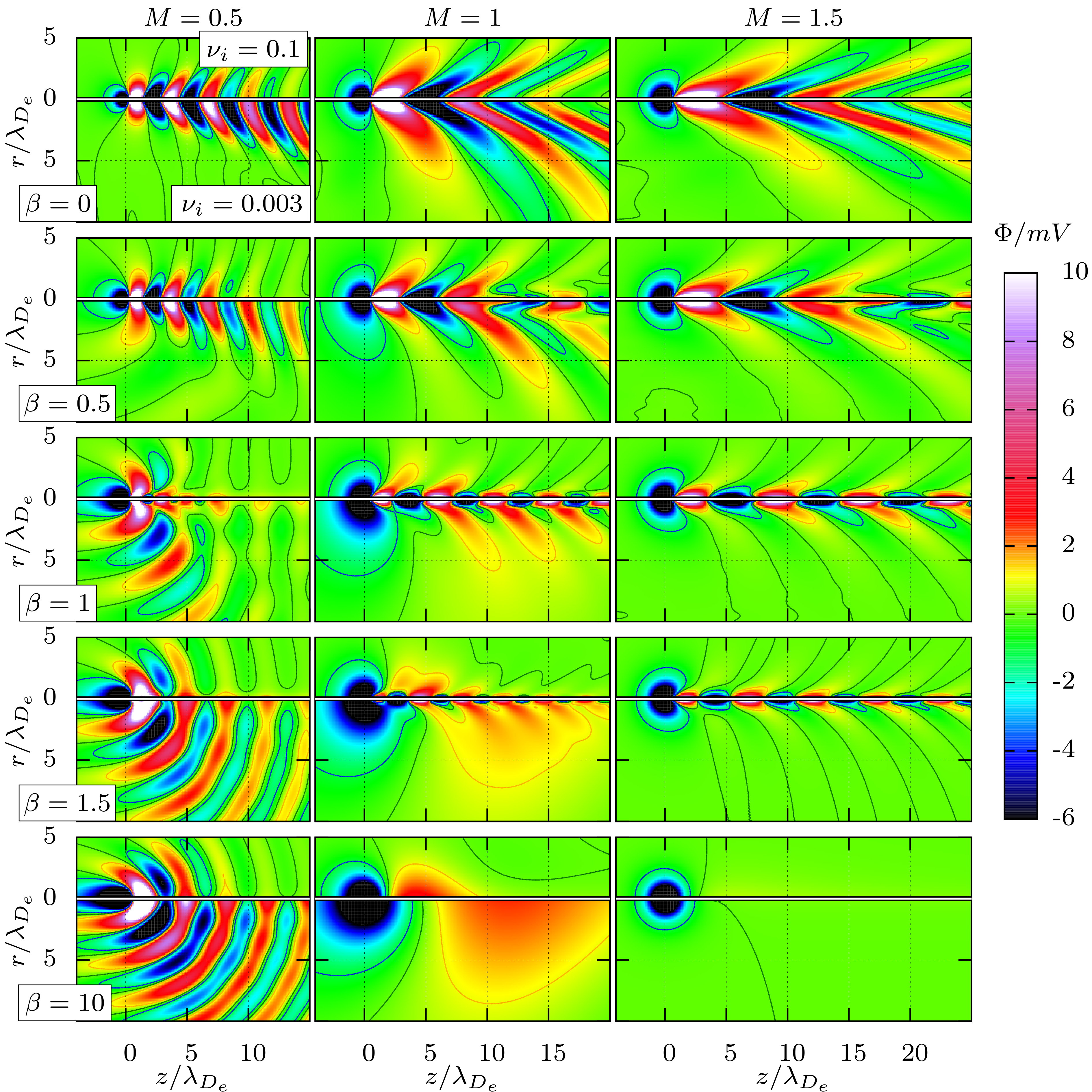}\hspace{2pc}%
\begin{minipage}[b]{\textwidth}\caption{\label{pic:matrix2}
Contour plot of plasma potential $\Phi(r)$ for  subsonic ($M=0.5$), sonic $M=1.0$, and supersonic ($M=1.5$) ion drift  (from left to right) given for five levels of the external magnetic field (increasing from top to bottom). 
For further settings see figure~\ref{pic:matrix1}.}
\end{minipage}
\end{figure}

So far, only the subsonic regime has been discussed. Figure~\ref{pic:matrix2} pictures a much broader range of $M$ values including $M=1$, and, as a representative supersonic streaming velocity, $M=1.5$. The subsonic case $M=0.5$ is shown for the sake of completeness and was already discussed in the context of figure~\ref{pic:matrix1}. 

At ion sound speed, corresponding to $M=1$, and $\beta=0$ the trend discussed before continues, i.e, the wake structure extends further since the ions carry four times more kinetic energy than for $M=0.5$, and their trajectories are far less affected by the grain. At finite magnetization, $\beta=0.5$, the effect of wave fronts being bent around the grain is no longer observed. Instead, at large distances from the grain, an irregular wake structure appears due to the disturbances by the external magnetic field. Increasing $\beta$ further suppresses the wake structure of the potential.  Compared to the unmagnetized limit, the lateral extension of the (negative) potential minima becomes substantially compressed. Furthermore, a symmetry breaking between positive and negative extrema becomes apparent.

At $\beta=1.5$, a positive contour develops that comprises several oscillations of the ion wake. Approaching very strong magnetization, $\beta=10$, the wake oscillations completely disappear, and only a single large positive potential area persists. While, in general, the topology of the wake structure does not depend on the strength of the collisional damping $\nu_i$, the position of this positive ion focus does crucially depend on $\nu_i$. Also, we note that in the direct vicinity of the grain, the (almost isotropic) screening of the Coulomb potential is very weak compared to the other cases considered so far. The wake structure as a whole appears as an (asymmetric) dipole-like entity.

Finally, let us consider the supersonic wake structure at $M=1.5$. In the unmagnetized case, the angle of the Mach cone is further reduced.
With an increasing external magnetic field, the oscillating wake structure becomes more and more compressed around the centre axis. The symmetry breaking between positive and negative extrema and the development of the enveloping structure as observed for $M=1.0$ is, however, not apparent in the supersonic regime. Instead of a dipole-like wake structure, one finds, in the limiting case of large values of $\beta$, an  isotropic grain potential without significant wake oscillations~\cite{nani2002,nina2003}. 

\begin{figure}
\includegraphics[width=0.935\textwidth]{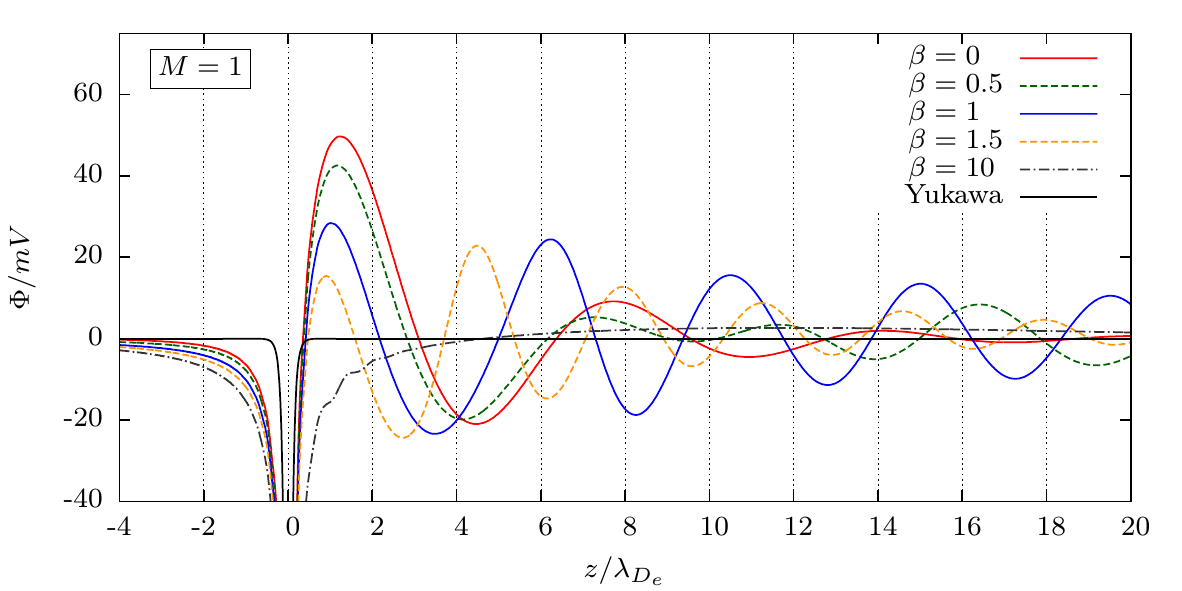}\hspace{2pc}%
\begin{minipage}[b]{\textwidth}\caption{\label{pic:axial1}
Same as figure \ref{pic:axial033} but for sonic ion drift $M=1$.}
\end{minipage}
\end{figure}

\begin{figure}
\includegraphics[width=0.935\textwidth]{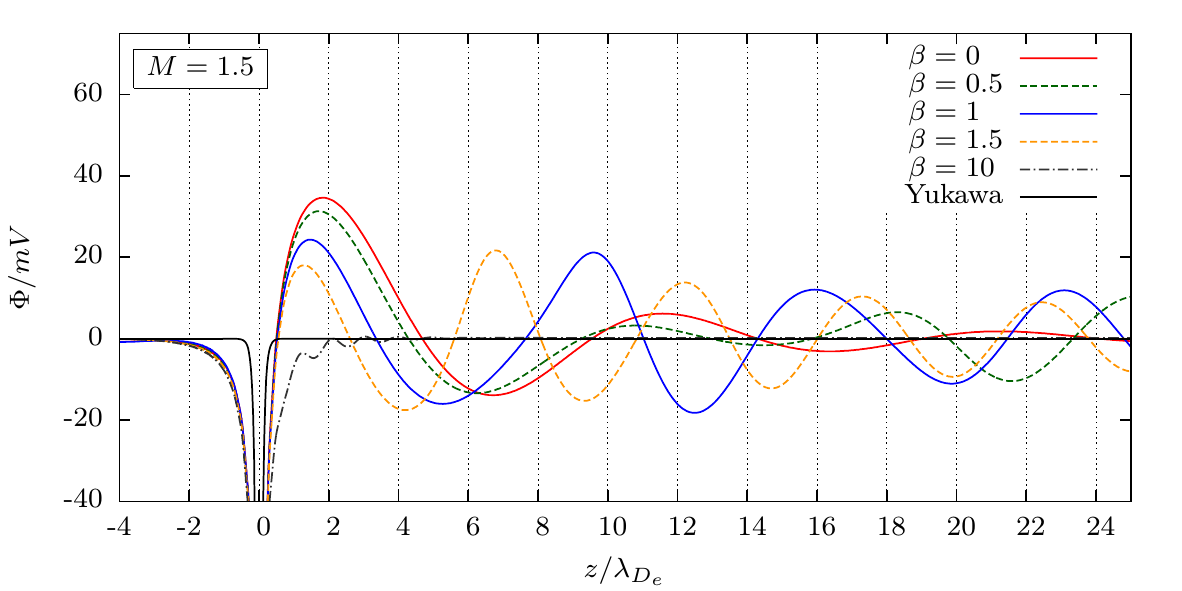}\hspace{2pc}%
\begin{minipage}[b]{\textwidth}\caption{\label{pic:axial15}
Same as figure \ref{pic:axial033} but for supersonic ion drift $M=1.5$.
Note that the wiggles in the range $(1-4)\lambda_{De}$ at $\beta=10$ are not a numerical artifact.

}
\end{minipage}
\end{figure}

The potential profile for sonic ion drift velocity, $M=1$, and $r=0$ is given in figure~\ref{pic:axial1}. In the upstream direction, screening is being reduced with increasing magnetization---similar to the subsonic regime. As $\beta$ is increased, the positions of the trailing peaks shift towards the grain.
In agreement with~\cite{nasa2001a, na2001,nani2002}, the amplitude of the trailing peak is found to be strongly damped compared to the unmagnetized case (see Fig.~\ref{pic:peak_height}).
Most interestingly, however, the exceptionally slow decay of the peak amplitudes for $\beta=1$ leads to far reaching dust-dust interactions. As mentioned above, for $\beta=10$, there exists an almost isotropically screened potential structure in the direct vicinity of the grain and only a single, far-reaching positively charged ion focus region downstream of the grain.

Considering the supersonic case $M=1.5$, figure~\ref{pic:axial15}, screening upstream of the grain is mostly independent of $\beta$, in contrast to the previously addressed cases. As observed for $M=0.66$ and $1.0$, the trailing peak shifts in the direction of the grain decreasing in amplitude. Again, the strongest peak amplitudes are observed for $\beta=1$.

\subsection{Characteristics of the wakefield extrema}

\begin{figure}[t]
\includegraphics[width=0.91\textwidth]{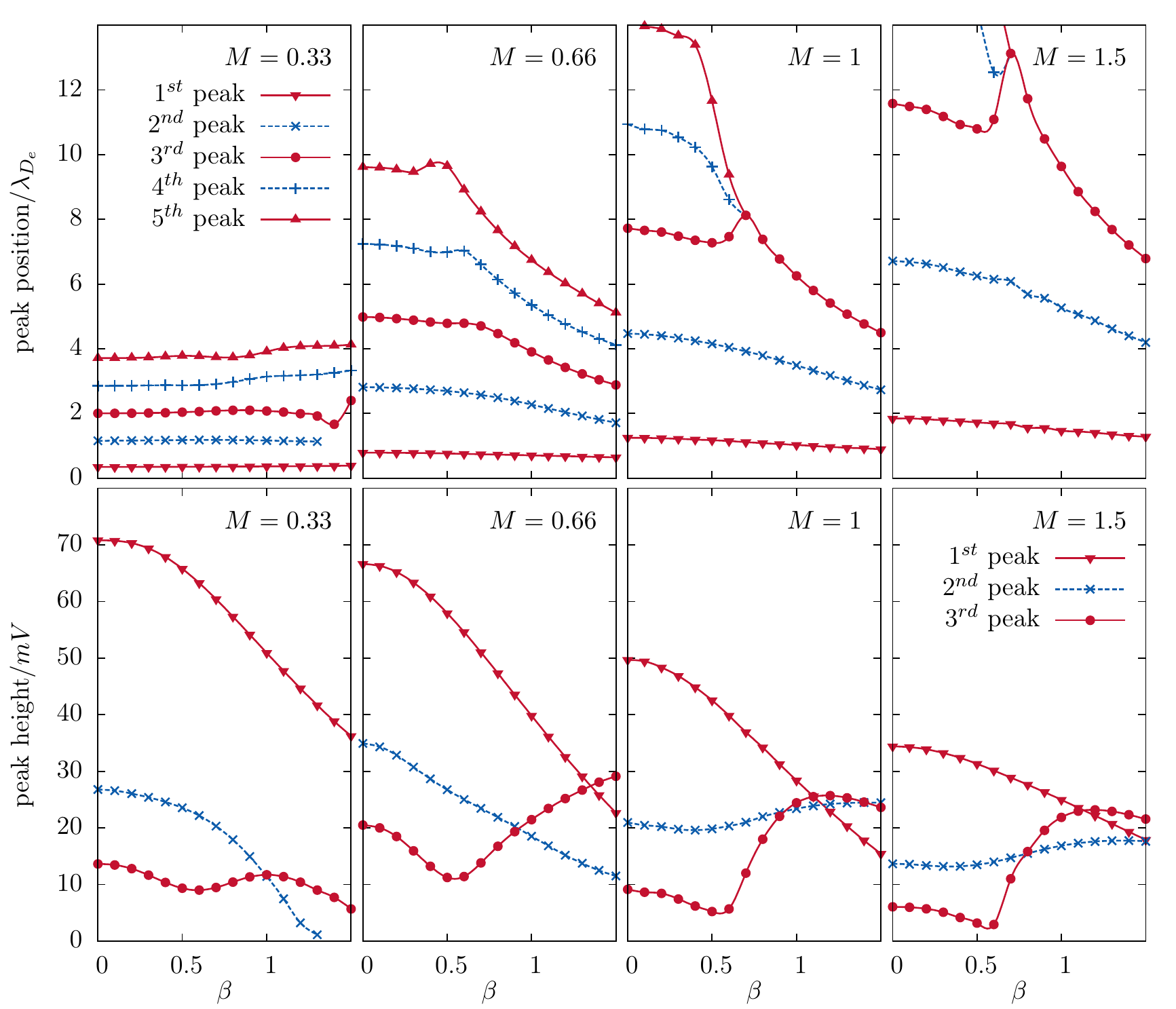}\hspace{2pc}%
\begin{minipage}[b]{\textwidth}\caption{\label{pic:peak_height}
Peak positions (top panels) and peak amplitudes (lower panels) of the trailing peaks along the $z$-axis as function of $\beta$ for four different drift velocities $M=0.33,0.66,1.0,1.5$.
Solid (red) lines indicate positive potential maxima, while crosses on dashed (blue) lines mark negative potential peaks. In total, for each of the four considered Mach numbers, results for $16\;\beta$-values, indicated by the symbols, have been evaluated on an up to $4096\times4096\times16384$ element sized grid.}
\end{minipage}
\end{figure}

The topology of the wake structure is essentially characterized by the position and the amplitude of the wakefield extrema along the $z$-axis. 
As seen in the contour plots, figure~\ref{pic:matrix1}, the amplitude of the off-axis extrema (in particular in the subsonic regime for $\beta\geq1$) is considerably lower than the dominant peak on the z-axis directly behind the grain. While the primary peaks have amplitudes far above  $20$mV, the off-centre peaks are well below $10$mV. Therefore, we restrict the following considerations to the on-axis wake extrema. 

As discussed before, with increasing $M$, the wake structure becomes more and more elongated, i.e., the distance of the individual peaks from the grain increases. This finding is well observed for any specific magnetization $\beta$, see figure \ref{pic:peak_height}. We note that in the unmagnetized case, $\beta=0$, the peak positions are equidistant~\cite{lumi2012}, while at finite magnetization deviations are observed. In turn, considering a fixed streaming velocity $M$, an increasing magnetization $\beta$ typically leads to a shift of the peak positions towards the grain. This observation is more pronounced for larger values of $M$. 

However, an anomaly to this general trend is found for the third extremum (the second positive maximum) for $M>0.66$ around $\beta\approx0.7$. The reason for this unusual behaviour lies in the fact that more distant peaks from the grain exhibit a larger gradient with respect to $\beta$, see the peak positions in the upper panel of figure~\ref{pic:peak_height}. This means that peaks may approach each other and finally merge, as can be seen for the representative example $M=1$. The third positive maximum approaches the second one creating a common plateau at $\beta=0.6$ (not shown), while the second minimum vanishes. Finally, the original second and third positive peaks overlap, creating an new broad wave crest with a particularly large peak amplitude at $\beta=0.7$.


The lower panel of figure \ref{pic:peak_height} shows the peak height as a function of $\beta$. Of highest relevance is the first trailing peak, where distinct peak heights (that may allow for particle attraction) are observed in the subsonic regime. In particular, for the considered parameters, the strongest peak is found for $\beta=0$ and $M=0.425$. As a general rule, above this value of $M$ the amplitude of the first peak decreases at constant $\beta$ with increasing $M$. Exceptions are found for strong magnetization $\beta>1$ and $M\gg1$.

Considering the functional dependence on $\beta$ at constant drift velocity, the first peak typically reduces with increasing magnetization monotonically. Higher order extrema show, however, a non-monotonic behaviour, where the third peak (second maximum) exhibits a minimum around $\beta=0.5$. Interestingly, for $M=0.66$ and $\beta>1.3$, the second maximum attains a larger potential height than the first (primary) peak. A similar effect is found for $M=1.0,1.5$ and $\beta\approx1.1$, see also the on-axis potential profiles in figures~\ref{pic:axial066},\ref{pic:axial1}, and \ref{pic:axial15}.



\subsection{Screening in perpendicular streaming direction}

\begin{figure}[t]
\includegraphics[width=0.935\textwidth]{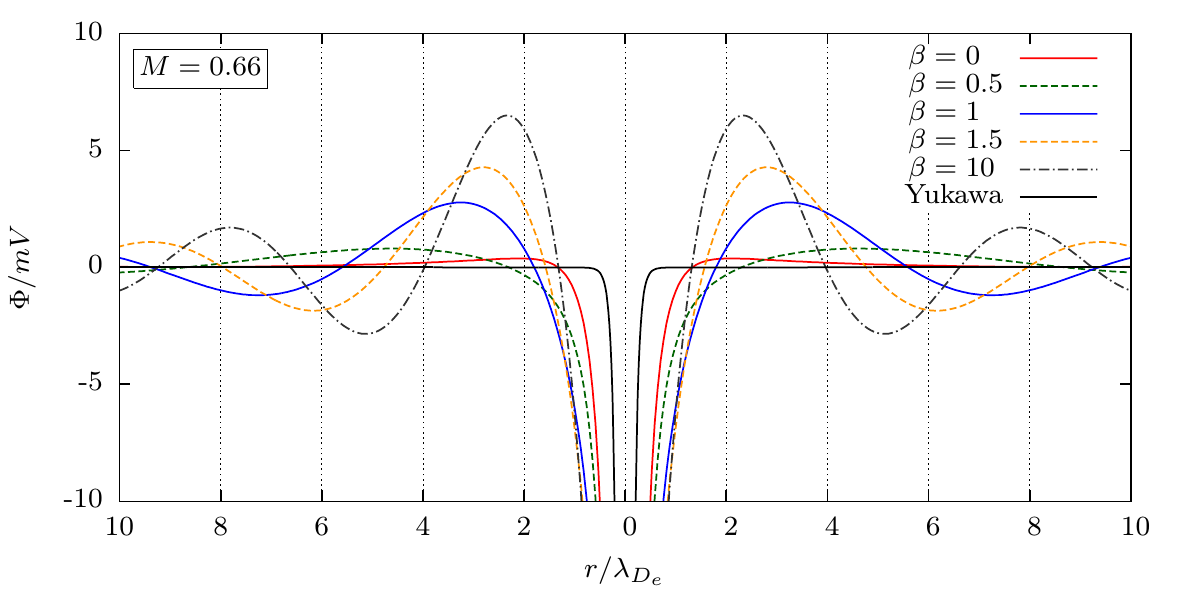}\hspace{2pc}%
\begin{minipage}[b]{\textwidth}\caption{\label{pic:axial_r_M066}
Potential cuts through the grain ($z = 0$) perpendicular to the ion flow direction for different magnetic inductions $\beta$ at $M=0.66$ ($\nu=0.003$, $T_e/T_i=100$). The Yukawa potential is shown for the corresponding static case $M=0$ (black solid line).}
\end{minipage}
\end{figure}

\begin{figure}[t]
\includegraphics[width=0.935\textwidth]{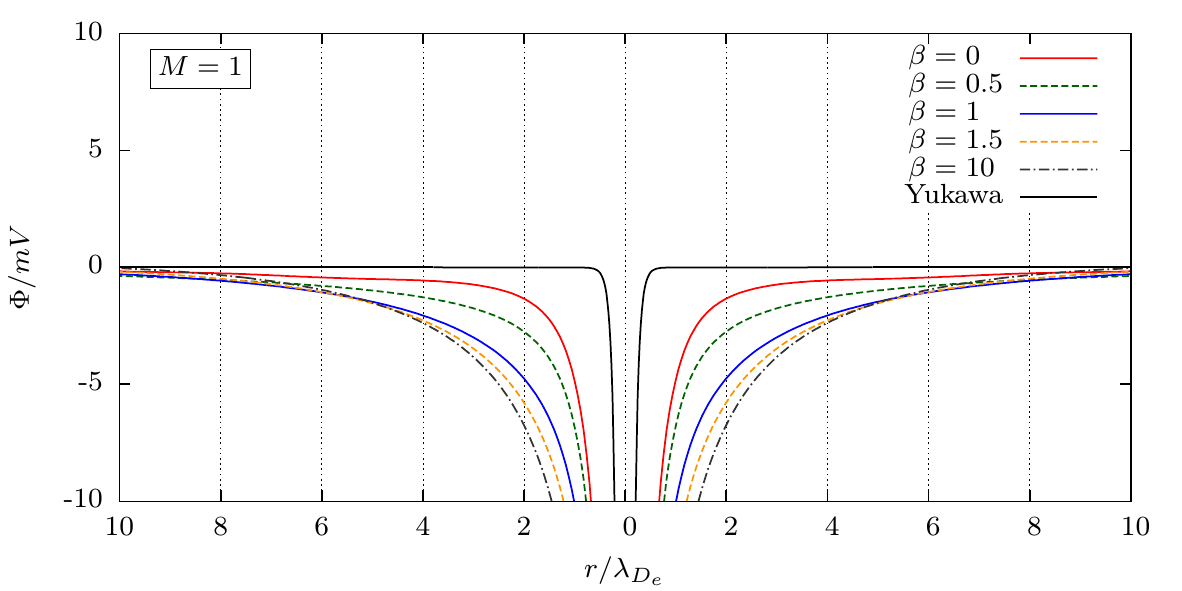}\hspace{2pc}%
\begin{minipage}[b]{\textwidth}\caption{\label{pic:axial_r_M1}
Same as figure \ref{pic:axial_r_M066} but for sonic ion drift $M=1$.}
\end{minipage}
\end{figure}

So far, we have discussed the potential profile along the streaming direction only. However, the question of wakefield oscillations in the perpendicular direction has also been under debate, see e.g,~\cite{komo2000,lumi2012,nani2002,nina2003}. In figure~\ref{pic:axial_r_M066}, we consider the effect of magnetization on the lateral potential profile for $M=0.66$. Without a magnetic field, $\beta=0$, a minor positive potential area is observed radially surrounding the grain. Under the influence of a finite magnetic field, the equipotential lines are bent around the grain, see figure~\ref{pic:matrix1}. This leads to strong oscillations of the potential radially outwards from the grain~\cite{nina2003}. These oscillations become stronger with increasing magnetization, but even when approaching very strong magnetization the peak amplitude is well below $10$\,mV.

At a slightly higher Mach number $M\geq1$, figure~\ref{pic:axial_r_M1}, a completely different picture is observed~\cite{nina2003}: The lateral wake oscillations completely vanish and the potential decay becomes strictly monotonic~\cite{nani2002}. In the vicinity of the grain $r/\lambda_{De}<5$, the effective shielding length perpendicular to the magnetic field is found to gradually increase with the magnetic induction.

\section{Conclusion}
We have presented a detailed analysis of the electrostatic potential of a charged dust grain in the presence of
a strong magnetic field in direction of the ion flow.
Our analysis is based on the kinetic theory result of the ion dielectric function with collisions included via a BGK collision term.

Our main focus was directed on the behavior of the dynamically screened grain potential when $M$ and $\beta$ are varied. 
It was shown that the effect of the magnetic field on the oscillatory wake structure strongly depends on the Mach number $M$. 
In the regime of subsonic ion flow, $M<1$, with increasing magnetization the equipotential lines are bent around the grain
with additional potential peaks appearing off the center axis and the amplitude of the wake maxima on the center axis being substantially reduced.
For supersonic ion flow velocities, $M>1$, the magnetic field radially compresses the plasma wakefield to the center axis, which completely vanishes in the limit of strong magnetization.

In recent experiments strong ion magnetization parallel to the direction of the ion flow  could be achieved~\cite{carst2012,2014jung}.
For two vertically aligned grains, a strong influence of the magnetic field on the ion-wake-mediated particle interaction
was observed when the magnetization of the ions exceeds $\beta\geq0.5$ (see Fig. 5(b) in \cite{carst2012}).
Analyzing the vertical coupling of the particle pair, a continuous reduction of the vertical grain attraction, i.e. damping of the ion focus, with an increasing magnetic field has been reported, which is in full accordance with our theoretical findings. 
For larger systems, our findings indicate that particles located adjacent to other grains are significantly more affected than in the unmagnetized case. 

Furthermore, our results should be of interest for dense quantum plasmas (warm dense matter), where wake effects are expected to exist for the 
ions~\cite{ludwigquantum,vlad2011qwakes,2014NJPzhandos}. Here, magnetization of streaming electrons should have similar effects as in the classical case.
Finally, we note that in this first systematic kinetic study of magnetization effects, 
we approximated the ion distribution function by a drifting Maxwellian. 
The effect of non-Maxwellian ions~\cite{schsch2005,ivzh2005,laro2012,huha2013} will be studied in a forthcoming work.



\ack

This work is supported by the Deutsche Forschungsgemeinschaft via SFB-TR 24 (project A7 and A9), the German Academic Exchange Service (DAAD), and the North-German Supercomputing Alliance (HLRN) via grant \texttt{shp00006}.

\section*{References} 


\end{document}